\documentclass[12pt]{article}

\usepackage{graphicx}
\begin{document}
\begin{center}
{\bf Nonlinear arcsin-electrodynamics }\\
\vspace{5mm}
 S. I. Kruglov
\footnote{E-mail: serguei.krouglov@utoronto.ca}
 \\

\vspace{5mm}
\textit{Department of Chemical and Physical Sciences, University of Toronto,\\
3359 Mississauga Road North, Mississauga, Ontario, Canada L5L 1C6}
\end{center}

\begin{abstract}

A new model of nonlinear electrodynamics with three parameters is suggested and
investigated. It is shown that if the external constant magnetic field is present the phenomenon of vacuum birefringence takes place. The indices of refraction for two polarizations of electromagnetic waves, parallel and perpendicular
to the magnetic induction field are calculated. The electric field of a point-like charge is not singular
at the origin and the static electric energy is finite. We have calculated the static electric energy of point-like particles for different parameters of the model. The canonical and symmetrical Belinfante energy-momentum tensors and dilatation current are obtained. We demonstrate that the dilatation symmetry and dual symmetry are broken in the model suggested.

\end{abstract}

\section{Introduction}

It is known that in Maxwell's electrodynamics a point-like charge has an infinite electromagnetic energy. This problem of singularity was solved in the Born-Infeld (BI) electrodynamics \cite{Born}, \cite{Infeld}, \cite{Plebanski} where a new parameter with the dimension of the length was introduced. The dimensional constant introduced in BI electrodynamics gives the upper bound on the possible electric field. Therefore, non-linear electrodynamics can give a finite electromagnetic energy of an electron. It should be noted that in QED one-loop quantum corrections give contributions to classical electrodynamics that result in the appearance of non-linear terms in the Lagrangian \cite{Heisenberg}, \cite{Schwinger}, \cite{Adler}. Some examples of non-linear electrodynamics were considered in \cite{Kruglov}, \cite{Kruglov1}, \cite{Kruglov2}, \cite{Kruglov3}, \cite{Gaete}, \cite{Hendi} and \cite{Kruglov5}. Thus, for strong electromagnetic fields in the vacuum we have to take into account non-linear effects. In addition, nonlinear electrodynamics can arise due to possible quantum gravity corrections.

In this paper we suggest and investigate a new model of nonlinear electrodynamics which has the same attractive feature as in BI electrodynamics: a finite electromagnetic energy of a point-like charge.
We introduce the dimensional parameters $\beta$, $\gamma$ and the dimensionless parameter $C$.
At $C=0$, $\gamma=0$ the model becomes Maxwell's electrodynamics. If $\gamma\neq 0$ the model possesses the phenomenon of birefringence. We note that in generalized BI electrodynamics \cite{Kruglov4} birefringence takes place.

The paper is organized as follows. In section 2, we postulate the Lagrangian of a new model of non-linear electrodynamics. The field equations are represented in the form of Maxwell's equations with
the electric permittivity $\varepsilon_{ij}$ and  magnetic permeability $\mu_{ij}$ tensors depending on electromagnetic fields. It is shown that the electric field of a point-like charge is not singular at the origin. In section 3 we investigate the effect of vacuum birefringence in the presence of a constant and uniform magnetic field.
The canonical and symmetrical Belinfante energy-momentum tensors, the dilatation
current and its non-zero divergence are obtained in  section 4. The section 5 is devoted to the conclusion.

We use the Heaviside-Lorentz system with $\hbar =c=\varepsilon_0=\mu_0=1$ , and the Euclidian metric. Greek
letters run from $1$ to $4$ and Latin letters run from $1$ to $3$.

\section{The model and field equations}

We postulate the Lagrangian density of nonlinear electrodynamics
\begin{equation}
{\cal L} = -{\cal F}-\frac{C}{\beta}\arcsin(\beta{\cal F})+\frac{\gamma}{2}{\cal G}^2,
 \label{1}
\end{equation}
where $\beta$, $\gamma$ are parameters with the dimension of length$^4$ ($\beta{\cal F}$, $\gamma{\cal G}$ are dimensionless) and $C$ is a dimensionless parameter. The Lorentz-invariants are ${\cal F}=(1/4)F_{\mu\nu}^2=(\textbf{B}^2-\textbf{E}^2)/2$, ${\cal G}=(1/4)F_{\mu\nu}\tilde{F}_{\mu\nu}=\textbf{E}\cdot \textbf{B}$, and $F_{\mu\nu}=\partial_\mu A_\nu-\partial_\nu A_\mu$ is the field strength tensor,  $\tilde{F}_{\mu\nu}=(1/2)\varepsilon_{\mu\nu\alpha\beta}F_{\alpha\beta}$ is a dual tensor ($\varepsilon_{1234}=-i$) and $A_\mu$ is the 4-vector-potential.
At $C = 0$, $\gamma=0$ the Lagrangian density (1) becomes the Maxwell Lagrangian density.
We expect that the non-linear model introduced is
an effective non-linear model of electrodynamics which occurs at strong electromagnetic fields.

From Eq. (1) and the Euler-Lagrange equations we obtain the equations of motion
\begin{equation}
\partial_\mu\left(F_{\mu\nu}+\frac{CF_{\mu\nu}}{\sqrt{1-\left(\beta{\cal F}\right)^2}}-\gamma {\cal G}\tilde{F}_{\mu\nu}\right)=0.
\label{2}
\end{equation}
With the help of the expression for the electric displacement field
$\textbf{D}=\partial{\cal L}/\partial \textbf{E}$ ($E_j=iF_{j4}$), we obtain
from Eq.(1)
\begin{equation}
\textbf{D}=\left(1+\frac{C}{\sqrt{1-\left(\beta{\cal F}\right)^2}}\right)\textbf{E}+\gamma{\cal G}\textbf{B}.
\label{3}
\end{equation}
Eq.(3) can be represented in the tensor form, $D_i=\varepsilon_{ij}E_j$, where the electric permittivity tensor $\varepsilon_{ij}$ is
\begin{equation}
\varepsilon_{ij}=\varepsilon\delta_{ij}+\gamma B_iB_j,~~~~\varepsilon=1+\frac{C}{\sqrt{1-\left(\beta{\cal F}\right)^2}}.
\label{4}
\end{equation}
The magnetic field is given by $\textbf{H}=-\partial{\cal L}/\partial \textbf{B}$ ($B_j=(1/2)\varepsilon_{jik}F_{ik}$, $\varepsilon_{123}=1$), and we obtain from Eq.(1)
\begin{equation}
\textbf{H}= \left(1+\frac{C}{\sqrt{1-\left(\beta{\cal F}\right)^2}}\right)\textbf{B}-\gamma{\cal G}\textbf{E}.
\label{5}
\end{equation}
Introducing the magnetic induction field $\textbf{B}_i=\mu_{ij}\textbf{H}_j$, one finds the inverse magnetic permeability tensor $(\mu^{-1})_{ij}$
\begin{equation}
(\mu^{-1})_{ij}=\varepsilon\delta_{ij}-\gamma E_iE_j.
\label{6}
\end{equation}
Field equations (2) can be rewritten, with the aid of Eqs. (3),(5), in the form of the first pair of Maxwell's equations
\begin{equation}
\nabla\cdot \textbf{D}= 0,~~~~ \frac{\partial\textbf{D}}{\partial
t}-\nabla\times\textbf{H}=0.
\label{7}
\end{equation}
The Bianchi identity
\begin{equation}
\partial_\mu \widetilde{F}_{\mu\nu}=0,
\label{8}
\end{equation}
gives the second pair of Maxwell's equations
\begin{equation}
\nabla\cdot \textbf{B}= 0,~~~~ \frac{\partial\textbf{B}}{\partial
t}+\nabla\times\textbf{E}=0.
\label{9}
\end{equation}
Equations (7),(9) are Maxwell's equations where the electric permittivity tensor $\varepsilon_{ij}$ and magnetic permeability tensor $\mu_{ij}$ depend on the fields $\textbf{E}$ and $\textbf{B}$.
Eqs. (3),(5) mimic a medium with complicated properties.
From equations (3),(5), one can obtain the equality $\textbf{D}\cdot\textbf{H}=(\varepsilon^2+2\gamma\varepsilon {\cal F}-\gamma^2{\cal G}^2)\textbf{E}\cdot\textbf{B}$. As $\textbf{D}\cdot\textbf{H}\neq\textbf{E}\cdot\textbf{B}$ the dual symmetry is broken \cite{Gibbons}. At $C=0$ ($\varepsilon=1$), $\gamma=0$, we arrive at classical electrodynamics and the dual symmetry is recovered. It should be noted that BI electrodynamics is dual symmetrical but in QED, due to one loop quantum corrections (Heisenberg-Euler Lagrangian), the dual symmetry is broken.

Let us consider electrostatics, $\textbf{B}=\textbf{H}=0$. Then the equation for the point-like charge is given by
\begin{equation}
\nabla\cdot \textbf{D}_0=e\delta(\textbf{r})
\label{10}
\end{equation}
with the solution
\begin{equation}
\textbf{D}_0=\frac{e}{4\pi r^3}\textbf{r}.
\label{11}
\end{equation}
Eq. (11), taking into consideration (4), becomes
\begin{equation}
E_0\left(1+\frac{C}{\sqrt{1-\beta^2 E_0^4/4}}\right)=\frac{e}{4\pi r^2}.
\label{12}
\end{equation}
The solution to Eq. (12) at $r\rightarrow 0$ is given by
\begin{equation}
E_0=\frac{\sqrt{2}}{\sqrt{\beta}}.
\label{13}
\end{equation}
Thus, the maximum electric field (13) has a finite value similar to BI electrodynamics. In linear electrodynamics the electric field strength possesses the singularity that results in an infinite electric energy of the point-like charged particle.

It is useful to define unitless values
\[
x=\frac{4\sqrt{2}\pi r^2}{e\sqrt{\beta}},~~~~y=\sqrt{\frac{\beta}{2}}E_0.
\]
Then Eq. (12) is rewritten as
\begin{equation}
x=\frac{\sqrt{1-y^4}}{y\left(\sqrt{1-y^4}+C\right)}.
\label{14}
\end{equation}
At $x\rightarrow 0$ ($r\rightarrow 0$), we have $y\rightarrow 1$ and this is equivalent to Eq. (13). When distance $r$ approaches to infinity, $x\rightarrow \infty$, and $y\rightarrow 0$.
As a result, the electric field of a point-like charged particle is finite at the origin.
The function $x(y)$ for different values of the parameter $C$ is presented in Fig.1.
\begin{figure}[h]
\includegraphics[height=3.0in,width=4in]{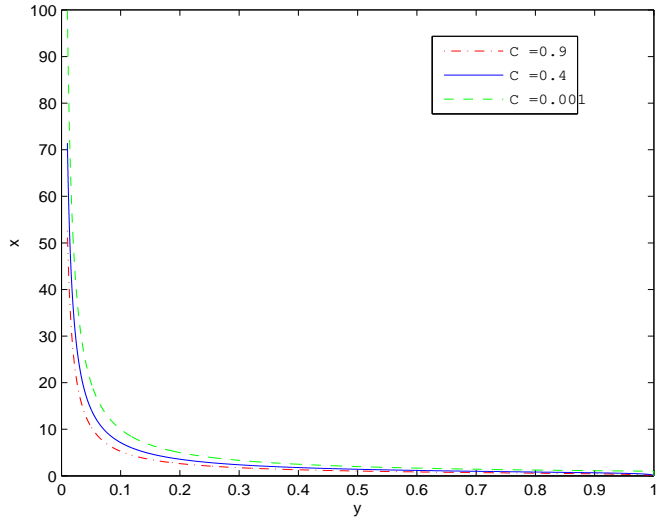}
\caption{\label{fig.1}$x$ versus $y$ for different values of the parameter $C$.}
\end{figure}

\section{Vacuum birefringence}

Vacuum birefringence within QED was investigated in \cite{Adler1}, \cite{Biswas}.
Let us consider the external constant and uniform magnetic induction field $\textbf{B}_0=(B_0,0,0)$ and the plane electromagnetic wave $(\textbf{e}, \textbf{b}$)
\begin{equation}
\textbf{e}=\textbf{e}_0\exp\left[-i\left(\omega t-kz\right)\right],~~~\textbf{b}=\textbf{b}_0\exp\left[-i\left(\omega t-kz\right)\right],
\label{15}
\end{equation}
propagating in the $z$-direction.
The total electromagnetic fields are $\textbf{E}=\textbf{e}$, $\textbf{B}=\textbf{b}+\textbf{B}_0$. We consider the case of the strong magnetic induction field. Then amplitudes of the electromagnetic wave, $e_0, b_0$, are small  compared to the magnetic induction field, and we have $e_0,b_0\ll B_0$. Linearizing Eq. (3),(5) one finds the electric permittivity and magnetic permeability tensors
\begin{equation}
\varepsilon_{ij}=\varepsilon\delta_{ij}+\gamma \delta_{i1}\delta_{j1}B_{0}^2,~~~\varepsilon=1+\frac{C}{\sqrt{1-\beta^2 B_0^4/4}},~~~\mu_{ij}=\mu\delta_{ij},~~~\mu=\varepsilon^{-1}.
\label{16}
\end{equation}
As $C\ll 1$ and $\beta^2 B_0^4\ll 1$, the term $C\beta^2 B_0^4$ in the expression for $\mu_{ij}$ was neglected.
From Maxwell's equations (7), (9) we obtain the wave equation
\begin{equation}
\partial_j^2 E_i-\mu\varepsilon_{ij}\partial^2_tE_j-\partial_i\partial_jE_j=0.
\label{17}
\end{equation}
In the case when the polarization is parallel to the external magnetic field, $\textbf{e}=e_0(1,0,0)$, and from Eq. (17) one finds the relation $\mu\varepsilon_{11}\omega^2=k^2$. Then the index of refraction is given by
\begin{equation}
n_\|=\sqrt{\mu\varepsilon_{11}}=\sqrt{1+\frac{\gamma B_0^2\sqrt{1-\left(\beta B_0^2\right)^2/4}}{C+\sqrt{1-\left(\beta B_0^2\right)^2/4}}}.
\label{18}
\end{equation}
If the polarization of the electromagnetic wave is perpendicular to the external induction magnetic field, $\textbf{e}=e_0(0,1,0)$, and the relation $\mu\varepsilon\omega^2=k^2$ holds. Then the index of refraction is
\begin{equation}
n_\perp=\sqrt{\varepsilon\mu}=1.
\label{19}
\end{equation}
As a result, the effect of vacuum birefringence takes place and the phase velocity depends on the polarization of the electromagnetic wave. When the polarization of the electromagnetic wave is parallel to the external magnetic field, $\textbf{e}_0\parallel \textbf{B}_0$, the speed becomes $v_\|=1/n_\|$ ($c=1$), and in the case $\textbf{e}\perp \textbf{B}_0$ the speed of the electromagnetic wave is $v_\perp=1$.

\section{The energy-momentum tensor and dilatation current}

From Eq. (1) and the expression of the canonical energy-momentum tensor
\[
T_{\mu\nu}^{c}=(\partial_\nu
A_\alpha)\frac{\partial{\cal L}}{\partial(\partial_\mu
A_\alpha)}-\delta_{\mu\nu}{\cal L},
\]
we obtain
\begin{equation}
T_{\mu\nu}^{c}=-(\partial_\nu A_\alpha)\left(F_{\mu\alpha}\varepsilon-\gamma{\cal G}\tilde{F}_{\mu\alpha}\right)-\delta_{\mu\nu}{\cal
L}. \label{20}
\end{equation}
The canonical energy-momentum tensor (20) is conserved,  $\partial_\mu T^c_{\mu\nu}=0$, but it is not
gauge-invariant and symmetrical tensor. To obtain the symmetric
Belinfante tensor we use the relation \cite{Coleman}:
\begin{equation}
T_{\mu\nu}^{B}=T_{\mu\nu}^{c}+\partial_\beta X_{\beta\mu\nu},
\label{21}
\end{equation}
where
\begin{equation}
X_{\beta\mu\nu}=\frac{1}{2}\left[\Pi_{\beta\sigma}\left(\Sigma_{\mu\nu}\right)_{\sigma\rho}
-\Pi_{\mu\sigma}\left(\Sigma_{\beta\nu}\right)_{\sigma\rho}-
\Pi_{\nu\sigma}\left(\Sigma_{\beta\mu}\right)_{\sigma\rho}\right]A_\rho,
\label{22}
\end{equation}
\begin{equation}
\Pi_{\mu\sigma}=\frac{\partial{\cal L}}{\partial(\partial_\mu
A_\sigma)}=-F_{\mu\sigma}\varepsilon+\gamma{\cal G}\tilde{F}_{\mu\sigma}.
\label{23}
\end{equation}
Because $X_{\beta\mu\nu}=-X_{\mu\beta\nu}$, the equation $\partial_\mu\partial_\beta X_{\beta\mu\nu}=0$ holds. As a result, the symmetrical Belinfante tensor is conserved, $\partial_\mu T^B_{\mu\nu}=\partial_\mu T^c_{\mu\nu}=0$. The generators of the Lorentz transformations $\Sigma_{\mu\alpha}$ have the matrix elements
\begin{equation}
\left(\Sigma_{\mu\alpha}\right)_{\sigma\rho}=\delta_{\mu\sigma}\delta_{\alpha\rho}
-\delta_{\alpha\sigma}\delta_{\mu\rho},
\label{24}
\end{equation}
and from Eqs. (22)-(24), we find
\begin{equation}
\partial_\beta X_{\beta\mu\nu}=\Pi_{\beta\mu}\partial_\beta A_\nu,
\label{25}
\end{equation}
and from Eq. (2) one obtains $\partial_\mu\Pi_{\mu\nu}=0$. Using
Eqs. (22),(25), we find the Belinfante tensor (21)
\begin{equation}
T_{\mu\nu}^{B}=-F_{\nu\alpha}\left(F_{\mu\alpha}\varepsilon-\gamma{\cal G}\tilde{F}_{\mu\alpha}\right)-\delta_{\mu\nu}{\cal L}.
\label{26}
\end{equation}
The trace of the energy-momentum tensor (26) is
\begin{equation}
T_{\mu\mu}^{B}=-\frac{4C{\cal F}}{\sqrt{1-(\beta{\cal F})^2}}+\frac{4C}{\beta}\arcsin(\beta{\cal F})+2\gamma{\cal G}^2.
\label{27}
\end{equation}
If $C=0$, $\gamma=0$ we arrive at Maxwell's electrodynamics, and the trace of the energy-momentum tensor (27) is zero. According to \cite{Coleman}, we use the modified dilatation current
\begin{equation}
D_{\mu}^{B}=x_\alpha T_{\mu\alpha}^{B}+V_\mu,
\label{28}
\end{equation}
and the field-virial $V_\mu$ is given by
\begin{equation}
V_\mu=\Pi_{\alpha\beta}\left[\delta_{\alpha\mu}\delta_{\beta\rho}
-\left(\Sigma_{\alpha\mu}\right)_{\beta\rho}\right]A_\rho.
\label{29}
\end{equation}
One can verify that $V_\mu=0$, and the modified dilatation current (28) is $D_{\mu}^{B}=x_\alpha T_{\mu\alpha}^{B}$.
The divergence of the dilatation current becomes
\begin{equation}
\partial_\mu D_{\mu}^{B}=T_{\mu\mu}^B.
\label{30}
\end{equation}
Thus, the dilatation (scale) symmetry is violated because the dimensional parameters $\beta$, $\gamma$ were introduced. It should be noted that in BI electrodynamics the scale and conformal symmetries are broken \cite{Kruglov4}, but linear Maxwell's electrodynamics is conformal symmetrical theory.

\subsection{Energy of the point-like charge}

Let us consider the total electric energy of the charged point-like particle, for example electron's energy.
One can obtain the energy density of the electric field ($\textbf{B}=0$) from Eq. (26)
\begin{equation}
\rho_E=T^B_{44}=E^2\left(\frac{1}{2}+\frac{C}{\sqrt{1-(\beta E^2)^2/4}}\right)-\frac{C}{\beta}\arcsin\left(\frac{\beta E^2}{2}\right).
\label{31}
\end{equation}
The variable $\beta^{1/4}{\cal E}$, where ${\cal E}=\int \rho_E dV$ is a total energy, becomes
\begin{equation}
\beta^{1/4}{\cal E}=\frac{e^{3/2}}{2^{11/4}\sqrt{\pi}}\int_0^\infty \sqrt{x}\left[\left(1+\frac{2C}{\sqrt{1-y^4}}\right)y^2-C\arcsin(y^2)\right]dx,
\label{32}
\end{equation}
where the function $x(y)$ is given by Eq. (14) ($e$ is the charge of the electron). Changing the variables in Eq. (32), one can represent (32) as
\[
\beta^{1/4}{\cal E}=\frac{e^{3/2}}{2^{11/4}\sqrt{\pi}}
\]
\begin{equation}
\times\int_0^1 \frac{\left[y^2\left(\sqrt{1-y^4}+2C\right)-C\sqrt{1-y^4}\arcsin(y^2)\right]\left[C\left(1+y^2\right)+(1-y^4)^{3/2}\right]}
{y^{5/2}(1-y^4)^{3/4}\left(\sqrt{1-y^4}+C\right)^{5/2}}dy.
\label{33}
\end{equation}
Numerical calculations of the integral (33) give the values $\beta^{1/4}{\cal E}$, for different parameters $C$, represented in Table 1.
\begin{table}[ht]
\caption{The normalized total energy $\beta^{1/4}{\cal E}$ of a point-like charged particle }
\vspace{5mm}
\centering
\begin{tabular}{c c c c c c c c c c c c}
\hline \hline 
$C$ & $0.00001$ & $0.0001$ & $0.001$ & $0.1$ & $0.2$ & $0.3$ & $0.4$ \\[0.5ex] 
\hline 
$\beta^{1/4}{\cal E}$ & $0.0271$ & $0.0277$ & $0.0279$ & 0.0263 & 0.0249 & 0.0238 & 0.0228\\
[1ex] 
\hline 
\end{tabular}
\label{table:crit}
\end{table}
If one assumes that the electron mass has a pure electromagnetic nature according to the Abraham and Lorentz idea \cite{Born1}, \cite{Rohrlich}, \cite{Spohn}, then ${\cal E}=0.51$~MeV. Taking the parameter $C$ to be $C=0.00001$, we obtain the length parameter $l_1=\beta^{1/4}=10.49$ fm. Thus, the electromagnetic nature of the electron may be realized in the model suggested.

\section{Conclusion}

Thus, we suggest a new model of nonlinear electrodynamics with three parameters $C$, $\beta$ and $\gamma$.  The values $\beta^{1/4}$, $\gamma^{1/4}$ possess the dimensions of the length and can be treated as affective constants due to some effects, for instance, quantum gravity (in electromagnetic units, according to Eq. (1), the constants $\beta^{-1/2}$ and $\gamma^{-1/2}$ have the same dimensions as the field strength). In the presence of the external constant and uniform induction magnetic field the phenomenon of vacuum birefringence takes place. In the particular case when $\gamma=0$ birefringence vanishes. The indices of refraction for two polarizations of electromagnetic waves, parallel and perpendicular to the magnetic field, were calculated. As a result, phase velocities of electromagnetic waves depend on polarizations if $\gamma\neq 0$. The canonical and symmetrical Belinfante energy-momentum tensors and the dilatation current were obtained showing that the dilatation symmetry is broken in the model considered. The scale symmetry is violated because the dimensional parameters $\beta$, $\gamma$ were introduced. We have demonstrated that the electric field of a point-like charge is not singular, and the maximum possible electric field is equal to $E_{max}=\sqrt{2}/\sqrt{\beta}$ similar to BI electrodynamics. We show that the static electric self-energy of point-like particles is finite, and therefore, one can speculate that the mass of the electron possesses a pure electromagnetic energy according to old idea.

\end{document}